\documentclass[12pt, a4paper]{article}

\usepackage{tikz-cd}
\usetikzlibrary{arrows,snakes,shapes.arrows,decorations.markings}
     \tikzset{>=triangle 90}
     \tikzstyle{bbc}=[draw,circle,fill=black,scale=.75]
     \tikzstyle{rc}=[circle,fill=red,scale=.6]
     \tikzstyle{wc}=[draw,circle,scale=.75]

\usepackage{color}

\usepackage{latexsym}
\usepackage{amsmath, mathdots, amscd, bm}
\usepackage{multirow}
\usepackage{amssymb}
\usepackage{graphicx}
\usepackage{euscript,slashed}
\usepackage{amsfonts}
\usepackage{verbatim}
\usepackage{epsfig}
\usepackage{hyperref}
\usepackage[margin=2cm,bottom=2.0cm]{geometry}
\usepackage{float}
\usepackage{cite}
\usepackage{framed}
\usepackage{empheq}
\usepackage{xcolor}
\usepackage{bbold}

\def\be{\begin{eqnarray}}
\def\ee{\end{eqnarray}}
\def\0{\nonumber}

\providecommand{\Tr}{\textnormal{Tr}}

\def\0{\nonumber}

\usepackage{makeidx}
\usepackage{cleveref}

\begin{document}

\begin{titlepage}
 
 
\begin{flushright}

\end{flushright}
 
\vskip 2cm
\begin{center}
 
{\LARGE\bf \boldmath On Instantons at Large Charge} 
 
 \vskip 2cm
 
{\large Andrea Cipriani and Raffaele Savelli}

 \vskip 0.9cm

  Dipartimento di Fisica \& INFN, Universit\`a di Roma ``Tor Vergata'', \\ Via della Ricerca
Scientifica 1, I-00133 Roma, Italy \\[2mm]

 \vskip 2cm
 
\abstract{\noindent The large R-charge limit of two-point functions of chiral primary operators in rank-one $\mathcal{N}=2$ superconformal field theories exhibits a universal behavior controlled by the effective field theory on their Coulomb branch. In the case of $SU(2)$ SQCD with four flavors, this behavior is expected to be independent of the exactly-marginal gauge coupling. We provide an analytic proof of this prediction by computing the correlators directly via supersymmetric localization. Our analysis clarifies the interplay between the weak-coupling expansion and the large-charge expansion, with special emphasis on the precise role played by gauge-theory instantons in the latter regime. We conclude with remarks on the implications of our results for analogous observables in Argyres-Douglas theories.}

\end{center}

\end{titlepage}

\tableofcontents

\section{Introduction}

It is often the case that observables in strongly-coupled quantum systems undergo remarkable simplifications in the limit of large quantum numbers, allowing for analytic treatments of otherwise inaccessible quantities. This includes large global charge and large spin (see e.g.~\cite{Gaume:2020bmp} and references therein) or even high energy and high particle number (see e.g.~\cite{Khoze:2018mey} and references therein).

One particularly well studied example is two-point functions of Coulomb-branch operators in four-dimensional $\mathcal{N}=2$ superconformal field theories (SCFTs) in the limit of large $U(1)_{\rm R}$ charge. In the case of rank-one theories, there is a universal formula governing the large-charge behavior of such correlators \cite{Hellerman:2017sur,Hellerman:2018xpi} (see also \cite{Hellerman:2020sqj,Hellerman:2021yqz,Hellerman:2021duh})
\be\label{Eq:GenHO}
\log\langle\mathcal{O}^n(0)\overline{\mathcal{O}}^n(\infty)\rangle &\underset{n \to \infty}{\simeq}& \log\Gamma(dn+\alpha+1)+\ldots\,,
\ee
where $d$ denotes the conformal dimension of the chiral-ring generator $\mathcal{O}$, $\alpha$ captures the difference between the $a$-anomaly of the SCFT and the one of a free vector multiplet, and the ellipses hide an affine dependence on $n$, with coefficients that are theory-dependent and do not admit a universal description. This formula was obtained mainly using a supersymmetric version of the effective field theory (EFT) developed in \cite{Hellerman:2015nra,Alvarez-Gaume:2016vff,Monin:2016jmo,Hellerman:2017veg}. For this reason, the equality \eqref{Eq:GenHO} holds up to terms that are non-perturbatively small in $n$, which signal the breakdown of the EFT.\footnote{See \cite{Hellerman:2021yqz} and especially \cite{Hellerman:2021duh} for an in-depth study of such terms.} The only control parameter used here is the large $n$, and the EFT analysis makes no reference to the possible existence of any other parameter, such as an exactly-marginal coupling.

In particular, the case $d=2$ corresponds to $SU(2)$ SQCD with four massless hypermultiplets in the fundamental representation, and constitutes the only rank-one interacting $\mathcal{N}=2$ SCFT admitting an exactly-marginal parameter, the complexified gauge coupling $\tau$. In general, two-point functions of half-BPS operators in this theory feature a highly non-trivial dependence on $\tau$, which satisfy well-known recursion relations \cite{Papadodimas:2009eu,Baggio:2014ioa,Baggio:2014sna}. But, remarkably and somewhat surprisingly, their profile at large charge is essentially independent of it at the perturbative level. More precisely, for large $n$, the $\tau$ dependence is either exponentially suppressed or entirely relegated to the affine (non-universal) part $A(\tau,\bar\tau)n+\tilde{B}(\tau,\bar\tau)$ of the asymptotic expansion \eqref{Eq:GenHO}.

Extremal correlators such as \eqref{Eq:GenHO} in gauge theory can be efficiently computed at weak coupling using the methods developed in \cite{Gerchkovitz:2016gxx} (see also \cite{Billo:2017glv}), which rely on the technique of supersymmetric localization \cite{Nekrasov:2002qd,Flume:2002az,Bruzzo:2002xf,Nekrasov:2003rj,Pestun:2007rz}.\footnote{Other references where two-point correlators of chiral/anti-chiral primary operators are computed using localization on the four-sphere include \cite{Baggio:2014ioa,Baggio:2014sna,Baggio:2015vxa,Baggio:2016skg,Rodriguez-Gomez:2016ijh,Rodriguez-Gomez:2016cem,Pini:2017ouj,Billo:2019job,Beccaria:2020hgy,Beccaria:2021hvt,Billo:2021rdb}. In $\mathcal{N}=4$ Super-Yang-Mills theory, extremal correlators of chiral primary operators are exact at tree level \cite{Bianchi:1999ge}.} This allowed the authors of \cite{Hellerman:2018xpi,Hellerman:2020sqj,Hellerman:2021yqz} to perform several high-accuracy cross-checks of their proposal \eqref{Eq:GenHO} for SQCD, based on numerical computations. The first analytic evidence for the EFT predictions was given in \cite{Grassi:2019txd}, where the authors showed that extremal correlators in $SU(2)$ SQCD admit a dual description in terms of matrix models, with the size of the matrices corresponding to the number of operator insertions.\footnote{The extension of this study to higher rank is challenging, due to the large degeneracy of heavy operators. Nevertheless, recent progress in this direction was achieved in \cite{Grassi:2024bwl} (see also \cite{Beccaria:2020azj,Brown:2025cbz}), laying the groundwork for a systematic treatment of large R-charge expansions in higher-rank SCFTs, where EFT predictions are still lacking.} However, for simplicity, they limited their analysis to the one-loop level in the gauge theory, neglecting the instanton sum. If the large-charge limit is performed at \emph{fixed} gauge coupling $g_{\rm YM}$, as the validity of the EFT requires, instantons are not parametrically suppressed.\footnote{Disregarding instantons is, instead, a justified approximation in a double-scaling limit $n\to\infty,g_{\rm YM}\to0$ with $\lambda\equiv ng^2_{\rm YM}$ fixed, which the authors of \cite{Grassi:2019txd} showed to exist for extremal correlators, as conjectured in \cite{Bourget:2018obm} (see also \cite{Beccaria:2018xxl}). Within the double-scaling limit, one can study the regime of large $\lambda$ to gain better control over exponential corrections to the EFT. However, as explained in \cite{Hellerman:2021duh}, although this regime shares features with the large-$n$ limit at fixed $g_{\rm YM}$, it inevitably washes out gauge-instanton corrections.} This calls for an extension of the study done in \cite{Grassi:2019txd} to a more complete analytic test of \eqref{Eq:GenHO} that includes the contribution of gauge-theory instantons.

The purpose of this paper is to fill this gap. Focusing on $SU(2)$ SQCD with four fundamental flavors, we will shed light on the interplay between the weak-coupling and the large-charge expansion of the two-point function of Coulomb-branch operators, by clarifying how the instanton sum precisely enters the profile of the correlator when the number of insertions is very large. Our strategy goes as follows. We will begin by working order by order in the instanton expansion, identifying patterns that persist as the instanton number increases. We then elevate these observations to a general argument valid for an arbitrary number of instantons. More concretely, we will argue that, at each instanton order, two clearly distinct contributions emerge in the partition function from the large-charge perspective: One that is polynomial in the expansion for large vacuum expectation values (vevs) of the vector-multiplet scalar and the other which is non-polynomial. This is somewhat natural because the large-charge limit pushes the scalar field far away from the origin. Although this intuition has some flaws at the one-loop level because, as demonstrated in \cite{Grassi:2019txd}, there are important effects at large charge originating from small vevs, it is perfectly realized at the instanton level. We will show that instantons not only leave the universal part of the correlator in \eqref{Eq:GenHO} unchanged, but also that the non-polynomial component of the instanton partition function enters the large-charge expansion only through exponentially-suppressed terms.

While we will limit an explicit account of the non-polynomial part to $2$ instantons, we will be able to analyze the polynomial part, and hence the large-vev regime, exactly in the gauge coupling. This enables us to derive closed-form expressions for the coefficients of the affine part of the asymptotic expansion \eqref{Eq:GenHO}. On the one hand, the coefficient of the term linear in $n$, $A(\tau,\bar\tau)$, was previously obtained in \cite{Hellerman:2020sqj} with different methods, which combine semiclassical computations with S-duality and recursion relations, and we find complete agreement with that result. On the other hand, the constant term, $\tilde{B}(\tau,\bar\tau)$, was discussed in \cite{Hellerman:2021yqz} through the decomposition $\tilde{B}(\tau,\bar\tau)=B(\tau,\bar\tau)-\log Z_{S^4}(\tau,\bar\tau)$, where $Z_{S^4}$ denotes the four-sphere partition function.
While $B$ and $Z_{S^4}$ are individually renormalization-scheme dependent, this dependence cancels in the combination $\tilde{B}$. In \cite{Hellerman:2021yqz}, only $B(\tau,\bar{\tau})$ was computed explicitly. Working up to fourth instanton order in the Pestun-Nekrasov scheme, we perfectly reproduce that result and show that it receives contributions exclusively from large vevs. Furthermore, restricting to the large-vev sector, we derive an all-instanton expression for $\tilde{B}(\tau,\bar{\tau})$, which, remarkably, turns out to be completely independent of $\tau$. This leads us to the scheme-independent conclusion that the entire $\tau$ dependence of $\tilde{B}$ originates from small vevs.

Although not directly related to our main results, we expect the methods and insights developed in this paper to shed light on the challenging problem of computing large-charge correlators via localization in other rank-one SCFTs, which are intrinsically strongly coupled \cite{Argyres:1995jj,Argyres:1995xn,Eguchi:1996vu,Minahan:1996cj}. This is particularly relevant for Argyres-Douglas theories, which arise at special loci of asymptotically-free gauge theories where the instanton tower is resummed. In these models, no supersymmetric recursion relations are known to govern the correlators, making direct tests of the EFT predictions at large charge especially important. While a more comprehensive study of this problem is left for future work, here we present a numerical determination of the large-vev contributions to the coefficients $A$ and $\tilde{B}$ for six celebrated rank-one non-Lagrangian theories, using a convenient normalization of the chiral-ring generator. The evidence gathered in this paper strongly suggests that the value obtained for $A$ is exact. By contrast, the numerical value of $\tilde{B}$ is expected to change once the small-vev contribution is taken into account.

The paper is organized as follows. We start in Section \ref{Sec:Review} with a lightning review of how extremal correlators are computed via localization and on their large-charge expansion. In Section \ref{Sec:RoleInst} we go to the core of our investigation with a fully general analysis of the first two instanton orders that clarifies their role in large-charge correlators. We proceed in Section \ref{Sec:AB4} by focusing on the regime of large vevs: First, we perform a computation up to four instantons to show the miraculous cancellations occurring in the large-charge expansion and to match the coefficients $A(\tau,\bar\tau)$ and $B(\tau,\bar\tau)$ to the existing literature; then, we analyze the large-vev sector exactly in the gauge coupling, showing that the scheme-independent coefficient $\tilde{B}$ does not depend on $\tau$. Finally, we draw our conclusions in Section \ref{Sec:Concl}, offering a speculative discussion about large-charge two-point functions in Argyres-Douglas SCFTs and a numerical computation of $A$ and $\tilde{B}$ for rank-one isolated models.

\section{The large-charge expansion}\label{Sec:Review}

This Section is devoted to setting up the stage for our investigation. We will be focusing on the only Lagrangian four-dimensional (interacting) $\mathcal{N}=2$ SCFT of rank one, namely $SU(2)$  SQCD with four massless fundamental hypermultiplets. This theory admits a one-dimensional conformal manifold, parametrized by the exactly-marginal (complexified) gauge coupling\footnote{In the following, we will always be considering the ``UV coupling'', i.e.~the parameter appearing in the Lagrangian, as opposed to the ``IR coupling'', i.e.~the complex-structure parameter of the Seiberg-Witten curve.}
\be
\tau&=&\frac{\theta}{2\pi}+\frac{4\pi {\rm i}}{g^2_{\rm YM}}\,.
\ee
The ring of chiral superconformal primary operators of this theory is also one-dimensional and is freely generated by
\be\label{Eq:CBop}
\mathcal{O}&=&-4\pi {\rm i}\,\Tr\phi^2\,,
\ee
where $\phi$ is the complex scalar in the vector multiplet. The vev of the above operator parametrizes the Coulomb branch of the space of vacua, where the $U(1)_{\rm R}$ symmetry is spontaneously broken. 

Central to the study of this theory in flat space are the two-point functions\footnote{We work in a basis in which the ring structure constants are trivialized: $\mathcal{O}^n(x)\mathcal{O}^m(0)\stackrel{x\to 0}{\longrightarrow} \mathcal{O}^{n+m}(0)$.}
\be\label{2ptF}
G_{2n}(\tau,\bar{\tau}) &=&|x|^{4n}\langle \mathcal{O}^n(x) \overline{\mathcal{O}}^n(0) \rangle_{\mathbb{R}^4}\,,
\ee
which determine the chiral-ring data and all the extremal correlators of the theory. Here, $n$ is a quarter of the $U(1)_{\rm R}$ charge of $\mathcal{O}^n$ (or equivalently a half of its conformal dimension), and for $n=0$ we have the identity operator (normalized so that $G_0=1$). Following \cite{Gerchkovitz:2016gxx}, these quantities are derived via a Gram-Schmidt orthogonalization procedure, starting from the following matrix of integrals
\be\label{eq:matrixInt}
    M^{(n)}_{i j}(\tau,\bar{\tau}) &=& \frac{1}{Z_{S^4}(\tau,\bar{\tau})} \,\partial_\tau^i \,\partial_{\bar{\tau}}^j \, Z_{S^4}(\tau,\bar{\tau}) \,, \qquad\quad i, j = 0, \dots, n\,,
\ee
where $Z_{S^4}(\tau,\bar{\tau})$ is Pestun's four-sphere partition function \cite{Pestun:2007rz}
\be\label{eq:Z_S4}
    Z_{S^4}(\tau,\bar{\tau}) &= &\int_\mathbb{R} \mathrm{d}a \,(2a)^2\, e^{-4 \pi\, \text{Im}\tau\, a^2}  \,\underbrace{\frac{H(2{\rm i}a)H(-2{\rm i}a)}{|H({\rm i}a)H(-{\rm i}a)|^4}}_{|Z_{\text{1L}}({\rm i} a)|^2} \, |Z_{\text{inst}}({\rm i}a, \tau)|^2\,.
\ee
The latter, in turn, is written in terms of Nekrasov's partition function (computed via localization) \cite{Nekrasov:2002qd,Flume:2002az,Bruzzo:2002xf,Nekrasov:2003rj}, whose one-loop part $Z_{\text{1L}}$ depends on the modulus $a$ (which is essentially the vev of $\phi$) through the Barnes-G function $H(x) \equiv G(1+x)G(1-x)$, but it is independent of the coupling. Instead, the instanton part $Z_{\text{inst}}$, calculated in the $\Omega$-background $\epsilon_1=\epsilon_2=1/R$ with $R$ the sphere radius,\footnote{We set $R=1$ throughout the paper to avoid cluttering the notation.} does depend on the coupling through a power-series expansion in the fugacity $q\equiv e^{2\pi{\rm i}\tau}$, which we will use momentarily. At fixed $n\geq1$, the two-point function \eqref{2ptF} is then obtained as the following ratio of determinants \cite{Gerchkovitz:2016gxx} (see also \cite{Billo:2017glv,Billo:2019job})
\be \label{eq:corr_def}
    G_{2n}(\tau,\bar{\tau}) &=& 16^n\, \frac{\det M^{(n)}(\tau,\bar{\tau})}{\det M^{(n-1)}(\tau,\bar{\tau})}\,.
\ee
Importantly, the $\tau$ dependence of such ratios has been shown to be governed by the following Toda-chain differential equation \cite{Papadodimas:2009eu,Baggio:2014ioa,Baggio:2014sna}
\be\label{Eq:Toda}
\frac{1}{16}\partial_\tau\partial_{\bar{\tau}}\log G_{2n}(\tau,\bar{\tau}) &=&\frac{G_{2n+2}}{G_{2n}}-\frac{G_{2n}}{G_{2n-2}}-G_2\,.
\ee

As it happens in several other instances in Quantum Field Theory in the presence of large quantum numbers, the quantity \eqref{eq:corr_def} undergoes a remarkable simplification when considering insertions with large R-charge (i.e.~large $n$); in particular, it can be reliably estimated without relying on the microscopic model, but rather using the EFT description of the Abelian vector multiplet on the Coulomb branch. That this should be the case can be readily inferred from a saddle-point approximation of the free energy $\mathcal{F}$, whereby one gets $\mathcal{F}\sim -2c\, a^2+4n\log(a)$,\footnote{$c$ is a coupling-dependent coefficient which, modulo numerical factors, reduces to $1/g^2_{\rm YM}$ at weak coupling.} the first term originating from the leading contribution in the large-radius expansion and the second one due to the operator insertions. Hence, at large $n$ but fixed Im$\tau$, $\phi$ typically acquires large vevs, $a\sim \sqrt{n/c}$, thus making both electrically- and magnetically-charged BPS particles very heavy. Building upon the EFT techniques developed in \cite{Hellerman:2015nra,Alvarez-Gaume:2016vff,Monin:2016jmo,Hellerman:2017veg}, the authors of \cite{Hellerman:2017sur,Hellerman:2018xpi} (see also \cite{Hellerman:2020sqj,Hellerman:2021yqz,Hellerman:2021duh}) were able to predict that the logarithm of $G_{2n}(\tau,\bar{\tau})$ has the following asymptotic expansion
\be\label{eq:log_corr}
    \log G_{2n}(\tau,\bar{\tau}) &\underset{n \to \infty}{\simeq}& (A(\tau,\bar{\tau})+ 4 \log2) \,n+\tilde{B}(\tau,\bar{\tau}) + \log\left[\Gamma\left(2 n + \frac{5}{2} \right)\right]\,,
\ee
which is understood to be valid up to exponentially-suppressed terms. Perhaps, the most striking feature of this formula is that, except for the linear and constant terms in $n$, the so-called $A$ and $\tilde{B}$ coefficients,\footnote{We stick with the notation of ref.~\cite{Hellerman:2021yqz}: $A$ depends on how the chiral-ring generator is normalized and $\tilde{B}$ on the normalization of the unit operator in a way that is independent of the renormalization scheme adopted.} the asymptotic expansion is completely independent of the gauge coupling. In other words, all the non-trivial dependence on $\tau$ of the two-point function is in $A, \tilde{B}$, and in the non-perturbative completion, which the EFT is insensitive to. This property stems from the absence of higher-derivative F-terms in the EFT on the Coulomb branch of rank-one theories, and was crucially used in \cite{Hellerman:2018xpi} to reduce $\eqref{Eq:Toda}$ to a set of \emph{algebraic} recursion relations. The latter were then exploited to fix the coefficients of all the negative powers of $n$, starting from the knowledge of the three terms $n\log n, \log n,1/n$, calculated explicitly with the EFT.

Recalling how such two-point functions are computed with localization, this coupling independence appears surprising, due to the presence of $\tau$ in $Z_{\rm inst}$, entering through the fugacity $q$. In fact, as was shown in \cite{Grassi:2019txd} using matrix-model techniques, if one simply ignores $Z_{\rm inst}$ in \eqref{eq:Z_S4} (setting it to $1$),\footnote{Note that, as remarked in \cite{Grassi:2019txd}, gauge-instanton contributions are \emph{not} parametrically suppressed when working at fixed Im$\tau$, as we do here. This is in contrast to the double-scaling limit also performed in \cite{Grassi:2019txd}, as well as in \cite{Hellerman:2018xpi}, where, however, electrically-charged BPS particles are no longer parametrically heavy, thereby threatening the validity of the EFT. See \cite{Hellerman:2021duh} for a detailed analysis of the difference between these two limits.} the computation via localization correctly reproduces the EFT prediction for the universal part of \eqref{eq:log_corr}. In what follows, we will prove that, inserting back $Z_{\rm inst}$, generates a coupling dependence in the correlator \eqref{2ptF} which \emph{only} affects linear, constant, and exponentially-suppressed terms of the large-$n$ expansion \eqref{eq:log_corr}. This provides the first analytic check of the EFT prediction.

\section{The role of gauge instantons}\label{Sec:RoleInst}

In this Section, we will scrutinize how exactly gauge-theory instantons affect the two-point function \eqref{2ptF} by performing a fully general analysis up to $2$ instantons. After introducing the relevant quantities in Section \ref{Sec:GenAn}, we first analyze the polynomial terms in the instanton partition function in Section \ref{Sec:pol}, and then go over the more intricate case of the non-polynomial remainder in Section \ref{Sec:rest}.

\subsection{Preliminaries}\label{Sec:GenAn}

The instanton part of the Nekrasov partition function appearing in \eqref{eq:Z_S4} reads explicitly like (see e.g.~\cite{Grassi:2019txd,Billo:2013fi,Fucito:2013fba})
\be\label{eq:InstExpZ}
Z_{\text{inst}} ({\rm i}a,\tau) &=& 1 + \frac{q}{2}\, (a^2-3) + \frac{q^2}{4}\, \frac{8 a^8+a^6-91 a^4-60 a^2+132}{(4 a^2+9)^2} + \mathcal{O}(q^3)
\ee
up to $2$ instantons. Note that, except for the one-instanton part, which is polynomial in $a$, all other components of $Z_{\text{inst}}$ are rational functions of $a$. Recalling the large-radius (or equivalently large-vev) expansion of the free energy $\mathcal{F}$, which is the logarithm of the partition function,\footnote{Crucially, all powers of $a$ higher than $2$ cancel out when computing the logarithm.}
\be\label{Eq:FE}
    \mathcal{F}(a,\tau) &=& \sum_{g=0}^\infty \mathcal{F}_g (\tau) \, a^{2-2g}\,,
\ee
we deduce that, at the one-instanton level, only $\mathcal{F}_0$ and $\mathcal{F}_1$ are non-vanishing. In view of this, we find it convenient to isolate the polynomial part\footnote{By ``polynomial part'' we simply mean all terms with non-negative powers of $a$ in the Taylor expansion of the instanton partition function around $a=\infty$. At the level of the free energy, this corresponds to isolating the instanton contributions to $\mathcal{F}_0$ and $\mathcal{F}_1$.} in $a$ and write \eqref{eq:InstExpZ} as
\be \label{eq:def_contributions}
    Z_{\text{inst}}({\rm i}a,\tau) &=& 1 + \frac{q}{2}\,(a^2-3) + \frac{q^2}{4}\,\underbrace{\left(\frac{a^4}{2}-\frac{35}{16} a^2 + \frac{13}{8}\right)}_{Z^{(2)}_{\text{pol}}(a)} + \frac{q^2}{4} \,\underbrace{\frac{3 (a^2+2)}{16 (4 a^2+9)^2}}_{Z^{(2)}_{\rm rest}(a)} + \,\mathcal{O}(q^3)\,.
\ee

Taking derivatives inside the integral sign, the matrix $M^{(n)}_{ij}(\tau,\bar\tau)$ in \eqref{eq:matrixInt} becomes 
\be\label{eq:M_ij}
   M^{(n)}_{i j}(\tau,\bar\tau) & = & \frac{1}{Z_{S^4}}  \int_\mathbb{R} \, \mathrm{d}a \, 4 a^2 \, |Z_{\text{1L}}({\rm i} a)|^2 \, \partial_\tau^i \partial_{\bar{\tau}}^j (e^{-4 \pi\, \text{Im}\tau\, a^2} |Z_{\text{inst}}({\rm i}a,\tau)|^2)\nonumber\\
   &=& \frac{1}{Z_{S^4}}  \int_\mathbb{R} \, \mathrm{d}a \, 4 a^2 \, |Z_{\text{1L}}({\rm i} a)|^2 \, (-1)^j(2\pi {\rm i})^{i+j} e^{-4 \pi\, \text{Im}\tau\, a^2} \\
    &&\hspace{-.7cm}\cdot  \left[a^{2i} + \frac{q}{2} (a^2+1)^i  (a^2-3) + \frac{q^2}{4} (a^2+2)^i \left(Z^{(2)}_{\text{pol}}(a)+Z^{(2)}_{\text{rest}}(a)\right)+\mathcal{O}(q^3)\right]\hspace{-.1cm} \cdot\hspace{-.1cm} \left[\begin{array}{c} i\to j \\ q \to \bar{q}\end{array}\right].\nonumber
\ee
In order to handle its determinant, we make use of a key device, the Andr\'eief identity, which relates the determinant of a matrix of integrals to a multivariate integral of determinants. As recalled in \cite{Grassi:2019txd}, given two sets of $n+1$ functions of one variable $\{f_k(y)\}_{k=0}^{n}, \{g_k(y)\}_{k=0}^{n}$ and a measure $\mathrm{d} \mu(y)$, this identity reads 
\be
   \underset{a \, b}{\det} \int \mathrm{d}\mu(y) f_a(y) g_b(y) &=& \frac{1}{(n+1)!}\,\int \prod_{j=0}^n \mathrm{d} \mu(y_j) \, \,\underset{a \, b}{\det} (f_a (y_b)) \, \,\underset{c \, d}{\det} (g_c (y_d))\,.
\ee
Applying it to the determinant of \eqref{eq:M_ij}, with $\mathrm{d}\mu(a) = \mathrm{d}a \, 4 a^2 \, |Z_{\text{1L}}({\rm i} a)|^2 \, e^{-4 \pi \text{Im}\tau \,a^2}$, we obtain
\be\label{eq:first_expr_det}
    \det M^{(n)}(\tau,\bar\tau) &=& \frac{(2\pi)^{n(n+1)}}{Z_{S^4}^{n+1}(n+1)!} \int_{\mathbb{R}^{n+1}} \prod_{k=0}^n  \mathrm{d}\mu(a_k) \,\prod_{i<j} (a^2_i-a^2_j)^2 \,|\det\mathcal{M}(\tau)|^2\,,
\ee
where we have defined the $(n+1)\times(n+1)$ matrix $\mathcal{M}(\tau)$ as a power series in $q$
\be\label{eq:detcalM}
\mathcal{M}(\tau)\,=\,\sum_{I=0}^\infty\,\left(\mathcal{Z}^{(0)}\right)^{-1}\mathcal{Z}^{(I)}\,\left(\frac{q}{2}\right)^I\,,&\;\; {\rm with}\;&\left\{\begin{array}{l}\mathcal{Z}^{(0)}_{ij} = (a^2_j)^i\\ \mathcal{Z}^{(1)}_{ij} = (a^2_j+1)^i (a^2_j-3)\\ \mathcal{Z}^{(2)}_{ij} = (a^2_j+2)^i \left(Z^{(2)}_{\text{pol}}(a_j)+Z^{(2)}_{\text{rest}}(a_j)\right)\\  \ldots\,, \end{array}\right.
\ee
and we have used that $\det\mathcal{Z}^{(0)}$ is the Vandermonde determinant. Defining $\mathcal{K}^{(I)}\equiv\left(\mathcal{Z}^{(0)}\right)^{-1}\mathcal{Z}^{(I)}$ and expanding the holomorphic determinant $\det\mathcal{M}(\tau)$ in \eqref{eq:first_expr_det} to the second order, we get
\be\label{Eq:HolM}
\det\mathcal{M}(\tau)&=&1+\frac{q}{2}\,\Tr\,\mathcal{K}^{(1)}+\frac{q^2}{4}\,\left[\Tr\left(\mathcal{K}^{(2)}-\frac{\left(\mathcal{K}^{(1)}\right)^2}{2}\right)+\frac{\left(\Tr\,\mathcal{K}^{(1)}\right)^2}{2}\right]+\mathcal{O}(q^3)\,,
\ee

In what follows, we are going to analyze the polynomial and non-polynomial contributions of the instanton partition function separately. We will find that the non-polynomial parts affect \emph{only} the non-perturbative completion of the large-$n$ expansion of the quantity $\log(Z_{S^4}G_{2n})$.

\subsection{Polynomial part}\label{Sec:pol}

We ignore for the moment the piece $Z^{(2)}_{\rm rest}(a_j)$ in \eqref{eq:detcalM}, and concentrate solely on the polynomial contributions. It is easy to check that for the coefficient of order $q$ we have
\be
\Tr\,\mathcal{K}^{(1)}&=&\frac{1}{2}(n+1)(n-6) + \sum_{\alpha=0}^n a^2_\alpha\,,
\ee
whereas, for the coefficient of order $q^2$ we have
\be
\Tr\left(\mathcal{K}^{(2)}|_{\rm pol}-\frac{\left(\mathcal{K}^{(1)}\right)^2}{2}\right)+\frac{\left(\Tr\,\mathcal{K}^{(1)}\right)^2}{2}&=&\frac{1}{16}\left(2 n^4 - 20 n^3 + 35 n^2 + 83 n + 26\right)\\
&&+\,\left(\frac{13}{16} + \frac{(n-6)(n+1)}{2}\right)\sum_{\alpha=0}^n a^2_\alpha + \frac{1}{2}\left(\sum_{\alpha=0}^n a^2_\alpha\right)^2.\nonumber
\ee

Performing the change of variables $a^2_k=x_k$ and integrating over half of the real lines, Eq.~\eqref{eq:first_expr_det} becomes
\be\label{eq:det_2inst__Mn}
   && \det M^{(n)}(\tau,\bar\tau) \hspace{.2cm} =\hspace{.2cm} \frac{4^{n+1}(2 \pi)^{n(n+1)}}{Z_{S^4}^{n+1}} \left[X^{(0)}_n + \frac{q+\bar{q}}{2} \, \left(X^{(1)}_n+\frac{(n+1)(n-6)}{2}\, X^{(0)}_n\right)  \right.  \nonumber \\
   && + \left. \frac{q^2+\bar{q}^2}{8} \left(X^{(2)}_n+\left(\frac{13}{8} + (n+1)(n-6)\right) X^{(1)}_n + \frac{2n^4-20 n^3+35 n^2+83 n+26}{8} \, X^{(0)}_n \right)  \right. \nonumber\\
     && +   \left. \frac{q \bar{q}}{4} \left(X^{(2)}_n+(n+1)(n-6)\, X^{(1)}_n + \frac{(n+1)^2(n-6)^2}{4}\, X^{(0)}_n\right) \right] +\mathcal{O}(q^3)\,, 
\ee
where we have defined
\be\label{eq:X_n}
     X^{(p)}_n &=& \frac{1}{(n+1)!} \int_0^\infty \prod_{k=0}^n \mathrm{d}x_k \, \sqrt{x_k} \, e^{-4 \pi \text{Im}\tau \, x_k} \, |Z_{\text{1L}}({\rm i} \sqrt{x_k})|^2 \prod_{i<j}(x_i-x_j)^2 \,\left(\sum_{\alpha=0}^n x_\alpha\right)^p\,. 
\ee

Before evaluating this quantity, recall that we are ultimately interested in the logarithm of the correlator \eqref{2ptF}. Therefore, taking the logarithm of \eqref{eq:det_2inst__Mn} and further expanding to the second order in $q$, we obtain
\be\label{Eq:detMCum}
\log \det M^{(n)} (\tau,\bar\tau) & =& \log\left(\frac{4^{n+1}(2 \pi)^{n(n+1)}}{Z_{S^4}^{n+1}}\right) + \log X_n^{(0)} \nonumber \\
&&+\, \frac{q+\bar{q}}{2} \left(\frac{(n+1)(n-6)}{2} +  \langle \Tr\,\chi\rangle_c\right) \nonumber \\
&&+\,\frac{q^2+\bar{q}^2}{64} \left((9n-46)(n+1) + 13\langle \Tr\,\chi\rangle_c \right) + \frac{(q+\bar{q})^2}{8} \langle (\Tr\,\chi)^2\rangle_c \nonumber\\
&&+\,\mathcal{O}(q^3)\,.
\ee
Above we have introduced a matrix $\chi$ with eigenvalues $\{x_\alpha\}_{\alpha=0,\ldots,n}$ and denoted by $\langle (\Tr\,\chi)^p\rangle$ the moment of $(\Tr\,\chi)^p$ with respect to the matrix ensemble defined by $X^{(0)}_n$ in \eqref{eq:X_n}, i.e.~$\langle (\Tr\,\chi)^p\rangle=X^{(p)}_n/X_n^{(0)}$. Note that the final result \eqref{Eq:detMCum} depends only on the so-called $p$-point \emph{cumulants} $\langle (\Tr\,\chi)^p\rangle_c$, that is, those core quantities in which $\langle (\Tr\,\chi)^p\rangle$ decomposes according to the partitions of the power $p$ \cite{Eynard:2015aea}. For instance, obviously $\langle \Tr\,\chi\rangle_c=\langle \Tr\,\chi\rangle$, whereas $\langle (\Tr\,\chi)^2\rangle_c=\langle (\Tr\,\chi)^2\rangle-\langle \Tr\,\chi\rangle^2$, and so on so forth. We will see, in Section \ref{Sec:AB4}, that this simple structure persists beyond $2$ instantons.

It is now time to compute the multivariate integral \eqref{eq:X_n} for a generic (non-negative) integer value of $p$. For $p=0$, this quantity has already been discussed in \cite{Grassi:2019txd}, where the authors wrote it in the following way
\be\label{Eq:X_nCG}
X^{(0)}_n=\frac{e^{12 (n+1) \log \gamma_G - (n+1) - \frac{(n+1)}{3} \log 2}}{t^{(n+1)(n+2)} \, (n+1)!} \int_0^\infty \prod_{k=0}^n \mathrm{d}z_k   e^{-z_k} \sqrt{z_k(z_k+t)}  U(z_k,t) \prod_{i<j}(z_i-z_j)^2.
\ee
Here, $\gamma_G$ is the Gleisher constant, $t\equiv 4 \pi \text{Im}\tau+8\log2$, and the $n+1$ variables have been changed to $z_k= t x_k$; moreover, $U(z,t)$ is a bounded function with the property that $U(z,t)\to1$ for $z\to \infty$, as follows directly from the large-$a$ asymptotics of $|Z_{\text{1L}}({\rm i} a)|^2$. The above multivariate integral can be studied by means of the Coulomb-gas formalism (see App.~A of \cite{Grassi:2019txd} and references therein). For the present purposes, the key upshot of this analysis is that the \emph{perturbative} large-$n$ expansion of the logarithm of the multivariate integral in \eqref{Eq:X_nCG} is \emph{independent} of $t$. More precisely, calling $N\equiv n+1$, one finds
\be\label{Eq:X_N}
\log X^{(0)}_{N-1}=N^2(\log N-\tfrac32)+N[\log(2^{2/3}\pi N\gamma_G^{12})-2]-N(N+1)\log t+\tfrac{7}{48}\log N+\mathcal{O}(1)\,.
\ee

Regarding the integral \eqref{eq:X_n} for $p>0$, we simply note that, since it is convergent, we can write it as\footnote{We thank Guillaume Dubach for suggesting this.}
\be\label{Eq:DerX0}
 X^{(p)}_n & = & (-1)^p\, \partial^p_t \, X^{(0)}_n\,,
\ee
which, using \eqref{Eq:X_N} and neglecting exponentially-suppressed terms at large $n$, means that
\be\label{eq:Yn_explicit_pert}
\langle (\Tr\,\chi)^p\rangle &\underset{n \to \infty}{\simeq}& \frac{[((n+1)(n+2)]_p}{t^p}\,,
\ee
where $[x]_p\equiv \prod_{i=0}^{p-1}(x+i)$ is the Pochhammer symbol.

Plugging \eqref{eq:Yn_explicit_pert} in \eqref{Eq:detMCum}, we get
\begin{align}\label{eq:final_log_detMn}
\log \det M^{(n)} (\tau,\bar\tau)  \underset{n \to \infty}{\simeq} &\log\left(\frac{4^{n+1}(2 \pi)^{n(n+1)}}{Z_{S^4}^{n+1}}\right) + \log X_n^{(0)} \nonumber \\
+& \frac{q+\bar{q}}{2} \left(\frac{(n+1)(n-6)}{2} +  \frac{(n+1)(n+2)}{t}\right) \nonumber \\
+&\frac{q^2+\bar{q}^2}{8} \left(\frac{(n+1)}{8t}\left(t(9n-46)+13(n+2)\right) \right) + \frac{(q+\bar{q})^2}{8} \frac{(n+1)(n+2)}{t^2}\nonumber\\
+&\,\mathcal{O}(q^3)\,,
\end{align}
where we see that, due to the fact that the moments appear only through their cumulants, all terms containing more than $2$ powers of $n$ drop out, in accordance with the EFT prediction.\footnote{In Section \ref{Sec:AB4}, by going beyond $2$ instantons, we will see that the moment $\langle (\text{Tr}\chi)^p\rangle$ is a polynomial in $n$ of degree $2p$, whereas the corresponding cumulant $\langle (\text{Tr}\chi)^p\rangle_c$ is only quadratic in $n$, independently of $p$. At the level of the present, order-by-order analysis, this appears as a miracle. But we will see in the all-order argument of Section \ref{Sec:AllOrd} that it has a clean geometric origin.}

Finally, using \eqref{eq:corr_def}, we obtain the logarithm of the two-point function:\footnote{To facilitate comparison to the existing literature, we find it convenient to give the instanton expansion of the sum $\log G_{2n}(\tau,\bar{\tau})+\log Z_{S^4}(\tau,\bar{\tau})$, thus removing the $n$-independent contribution of the $S^4$ partition function. In this way, the coefficient of $n^0$ will be $B(\tau,\bar{\tau})$ and not $\tilde{B}(\tau,\bar{\tau})$.}
\be\label{eq:final_logG2n}
\log(Z_{S^4}(\tau,\bar{\tau}) G_{2n}(\tau,\bar{\tau})) &\underset{n \to \infty}{\simeq}& \log(Z_{S^4}(\tau,\bar{\tau}) G_{2n}(\tau,\bar{\tau}))_0\nonumber\\
&&+\, \frac{q+\bar{q}}{2} \left(n-3 +  \frac{2(n+1)}{t}\right) \nonumber \\
&&+\,\frac{q^2+\bar{q}^2}{32} \left(9 n - 23+\frac{13(n+1)}{t}\right)+ \frac{(q+\bar{q})^2}{4} \,\frac{n+1}{t^2}\nonumber\\
&&+\,\mathcal{O}(q^3)\,,
\ee
where we have isolated the $0$-instanton contribution
\be\label{eq:0instlogG}
\log(Z_{S^4}(\tau,\bar{\tau}) G_{2n}(\tau,\bar{\tau}))_0 &\underset{n \to \infty}{\simeq}& 2n\left(\log\frac{8\pi}{t}-1\right)+\log\frac{2^{8/3}\pi\gamma_G^{12}}{t^2}-1+(2n+2)\log n\nonumber\\ && +\frac{47}{48n}+\mathcal{O}\left(\frac{1}{n^2}\right).
\ee
As can be verified, the coefficients of $n\log n, \log n,1/n$ coincide with the corresponding ones of $\log\Gamma(2n+\tfrac52)$ in \eqref{eq:log_corr}, whereas the instanton corrections in \eqref{eq:final_logG2n} only affect the coefficients $A(\tau,\bar{\tau})$ and $B(\tau,\bar{\tau})$ of the perturbative large-$n$ expansion, i.e.~the terms of order $n^1$ and $n^0$, respectively.\footnote{We remark that the non-perturbative terms, which we are neglecting throughout our study, are expected to receive instanton corrections, affecting their $\tau$ dependence.}

\subsection{Non-polynomial part}\label{Sec:rest}

We now pass to examine the non-polynomial piece $Z^{(2)}_{\rm rest}(a_j)$ in \eqref{eq:detcalM}, defined in \eqref{eq:def_contributions}, which we have been ignoring so far.
At the instanton order we are working, this gives contributions that simply add up to the final result for $\log \det M^{(n)}$ found in Section \ref{Sec:pol} (Eq.~\eqref{eq:final_log_detMn}).

The first step is to use the fact that $Z^{(2)}_{\rm rest}(a)$ is an analytic function to rewrite it as a geometric series around the origin:
\be
Z^{(2)}_{\rm rest}(a)&=&\frac{a^2+2}{432}\,\sum_{k=0}^\infty \,\left(-1\right)^k\,\sum_{s=0}^k \,
\begin{pmatrix}
k \\
s
\end{pmatrix}
 \frac{2^{4k-s}}{3^{4k-2s}} (a^2)^{2k- s}\,.
\ee
Hence, setting $r=2k-s$, the non-polynomial part of the matrix $\mathcal{Z}^{(2)}$ in \eqref{eq:detcalM} has the following expansion
\be
\mathcal{Z}^{(2)}_{ij}|_{\rm rest}&=& \frac{(a_j^2+2)^{i+1}}{432}\sum_{r=0}^{\infty}\,c_r \,(a_j^2)^r\,,
\ee
with coefficients given by
\be
c_r &=& \sum_{k=\lfloor \frac{r+1}{2} \rfloor}^r (-1)^k 
    \begin{pmatrix}
        k \\
        2k-r
\end{pmatrix}
    \frac{2^{2k+r}}{3^{2r}}\,.
\ee

Now we can study the $n$-dependence of the non-polynomial part of $\Tr\,\mathcal{K}^{(2)}$ appearing in \eqref{Eq:HolM} order by order in $r$. For example, it is easy to check that one has
\be
 432\,\Tr\left(\left(\mathcal{Z}^{(0)}\right)^{-1}\mathcal{Z}^{(2)}|_{\rm rest}\right)=\left\{\begin{array}{ll}(n+2)(n+1) + \sum_\alpha a^2_\alpha&\; r=0\\ \\  -\frac{8}{9}\left[\frac{2}{3}(n+2)(n+1) n + 2(n+1) \sum_\alpha a^2_\alpha + \sum_\alpha a^4_\alpha\right]&\; r=1\\ \\ \begin{array}{l}\frac{16}{27}\left[\frac{1}{3}(n+2)(n+1) n(n-1)+2(n+1)n\sum_\alpha a^2_\alpha\right. \\ \left. \quad\; + 2(n+1) \sum_\alpha a^4_\alpha +2\sum_{\alpha<\beta}a_\alpha^2a_\beta^2+\sum_\alpha a^6_\alpha \right]\end{array} &\; r=2 \\ \\ \hspace{4.5cm} \vdots & r\geq3\,. \end{array}\right.\nonumber
\ee
As one can see, at each value of $r$ there appear different functions of $n$ multiplying different polynomials of the $a_\alpha^2$'s. We name them $f_r^{(p_\ell)}(n)$, where $\{p_\ell\}_\ell$ label all the different functions of $n$ that multiply the various polynomials of degree $p$ showing up at level $r$. For example: There is obviously a single polynomial of degree $0$, $P_0(a_\alpha^2)=1$, and $f_0^{0}(n)=(n+2)(n+1), f_1^{0}(n)=\tfrac23(n+2)(n+1)n$, and so on; there is still a single polynomial of degree 1, $P_1(a_\alpha^2)=\sum_\alpha a_\alpha^2$, and $f_0^{1}(n)=1, f_1^{1}(n)=2(n+1)$, and so on. There are, instead, two distinct polynomials of degree $2$, the sum of squares of the $a^2_\alpha$'s, $P_{2_1}(a_\alpha^2)=\sum_\alpha a_\alpha^4$, and the symmetric degree-$2$ polynomial of the $a^2_\alpha$'s, $P_{2_2}(a_\alpha^2)=\sum_{\alpha<\beta} a_\alpha^2 a_\beta^2$: Here we have $f_0^{(2_1)}(n)=0,f_1^{(2_1)}(n)=1,f_2^{(2_1)}(n)=2(n+1)$, and so on, whereas $f_0^{(2_2)}(n)=f_1^{(2_2)}(n)=0,f_2^{(2_2)}(n)=2$, and so on. We can easily obtain a closed-form expression of all such functions for generic $r$. Here we list the first few of them (up to the quadratic order):\footnote{The functions $f_r^{(p_\ell)}(n)$ not explicitly written are meant to be zero.}
\be
f_r^{(0)}(n)&=&2^{r+1}\begin{pmatrix}
        n+2 \\
        r+2
\end{pmatrix} \qquad 0\leq r\leq n \,,\\
f_r^{(1)}(n)&=& 2^r\begin{pmatrix}
        n+1 \\
        r
\end{pmatrix}\qquad\quad 0\leq r\leq n+1\,, \\
f_r^{(2_1)}(n)&=& 2^{r-1}\begin{pmatrix}
        n+1 \\
        r-1
\end{pmatrix}\qquad 1\leq r\leq n+2\,, \\
f_r^{(2_2)}(n)&=& 2^{r-1}\begin{pmatrix}
        n \\
        r-2
\end{pmatrix}\qquad 2\leq r\leq n+2\,.
\ee

At this point, it is useful to compactly write the entire non-polynomial contribution to \eqref{Eq:HolM} as follows
\be
\frac{q^2}{4}\,\Tr\,\mathcal{K}^{(2)}|_{\rm rest}&=&\frac{q^2}{1728}\,\sum_{\{p_\ell\}}f^{(p_\ell)}(n)\,P_{p_\ell}(a_j^2)\,,
\ee
where we have defined $f^{(p_\ell)}(n)=\sum_r c_r\,f_r^{(p_\ell)}(n)$, and $P_{p_\ell}(a_j^2)$ are degree-$p$ polynomials in the variables $\{a_\alpha\}_{\alpha=0,\ldots,n}$ invariant under the permutation group $S_{n+1}$. The multivariate integral replaces such polynomials with the corresponding moments with respect to the matrix ensemble defined by \eqref{eq:X_n}. Therefore, the non-polynomial contribution to the logarithm of $\det M^{(n)} (\tau,\bar\tau)$, to be added to \eqref{eq:final_log_detMn}, can finally be written as
\be\label{Eq:logsetMrest1}
\log\det M^{(n)} (\tau,\bar\tau)|_{\rm rest}&\simeq&\frac{q^2+\bar{q}^2}{1728}\,\sum_{p=0}^\infty F^{(p)}(n)\,t^{-p}\,,
\ee
where we have used that
\be
\langle P_{p_\ell}(a_j^2)\rangle &\simeq& t^{-p}\, P_{p_\ell}(n)\,,
\ee
with $P_{p_\ell}(n)$ certain polynomials in $n$ whose explicit form is irrelevant for the present purposes,\footnote{For example, $P_0(n)=1$ while $P_1(n)=(n+1)(n+2)$ (see \eqref{eq:Yn_explicit_pert}). Strictly speaking, the higher ones are polynomials if we work with the Wishart-Laguerre ensemble emerging from the large-eigenvalue limit of $|Z_{\text{1L}}({\rm i} a)|^2$ \cite{Grassi:2019txd,Livan:2011uvs}. Nevertheless, we argue that, in the ensemble defined by \eqref{eq:X_n}, they all grow no faster than a power for large $n$.} and $F^{(p)}(n)=\sum_\ell f^{(p_\ell)}(n) P_{p_\ell}(n)$. 

The last step is to determine the large-$n$ behavior of the functions $F^{(p)}(n)$. Let us do it for the first few of them, assuming for the sake of simplicity that $n$ is even (for odd $n$ the same conclusions apply). For $F^{(0)}(n)$ we obtain 
\be
    F^{(0)}(n) \,=\, \sum_{r=0}^n c_r \, f_r^{(0)}(n)\,  = \,\frac{81}{32} - \frac{1}{32} (8n+17) \, e^{-2n\,\log3}\,,
\ee
where we recognize that $F^{(0)}(n)$ tends to a constant exponentially fast. Recalling that the constant disappears when computing the logarithm of the correlator using \eqref{eq:corr_def}, we indeed verified that this piece does not contribute to the asymptotic expansion of $\log(Z_{S^4}G_{2n})$.

As for $F^{(1)}(n)$, we find 
\be
    F^{(1)}(n) \,= \, (n+1)(n+2)\,\sum_{r=0}^{n+1} c_r \, f_r^{(1)}(n)\, =\, - \frac{1}{9}(n+1)(n+2)(8n+7)\,e^{-2n\,\log3}\,,
\ee
which shows that also this piece yields only non-perturbative corrections to the logarithm of the ($Z_{S^4}$-rescaled) two-point function.

Finally, $F^{(2)}(n)$ contains two contributions:
\be
F^{(2)}(n)&=&P_{2_1}(n) \,\sum_{r=1}^{n+2} c_r \, f^{(2_1)}_r(n)+P_{2_2}(n) \,\sum_{r=2}^{n+2} c_r \, f^{(2_2)}_r(n)\nonumber\\
&=&\left(\frac{8}{81}(4n+3)P_{2_1}(n)-\frac{32}{81}(8n-3)P_{2_2}(n)\right)\,e^{-2n\,\log3}\,,
\ee
which again display the same exponentially-suppressed behavior. Since the structure of the functions $f_r^{(p_l)}(n)$ is always the same, we expect that this behavior will not change beyond the quadratic order in $1/t$, although we have not verified it explicitly.\\

To summarize, we have shown that the non-polynomial part of the $2$-instanton correction contributes only exponentially suppressed terms in the large-$n$ expansion of the logarithm of the ($Z_{S^4}$-rescaled) correlator. Therefore, up to $2$ instantons, its perturbative asymptotic expansion is fully captured by Eq.~\eqref{eq:final_logG2n}, i.e.~by the polynomial parts of the instanton corrections, or equivalently by the large-vev regime of the free energy.

The procedure we have outlined can be equally well applied beyond $2$ instantons, separating the polynomial part of the corrections from the non-polynomial remainder. We argue that, by doing so, one will arrive at the same conclusion. Motivated by this conjecture, in the next Section we will first push the analysis up to $4$ instantons, but focusing on the polynomial parts only: We will compute the large-$n$ expansion of the logarithm of the ($Z_{S^4}$-rescaled) two-point function and find perfect match with earlier results in the literature. This provides non-trivial support to our claim that $\mathcal{F}_0$ and $\mathcal{F}_1$ alone fully encode the large-charge expansion at the perturbative level. Armed with this evidence, we will perform an all-instanton computation in the large-vev regime. Besides confirming our claim, this will provide an independent exact derivation of the coefficient $A$ and, at the same time, the first explicit derivation of the scheme-independent coefficient $\tilde{B}$,\footnote{We recall that in \cite{Hellerman:2021yqz} $\tilde{B}$ is expressed as $\tilde{B}=B-\log Z_{S^4}$ and, by specifying a renormalization scheme for the partition function, only $B$ is computed, whereas $\log Z_{S^4}$ is left implicit.} albeit only through its large-vev contribution.

\section{The large-vev sector}\label{Sec:AB4}

In this Section we will concentrate on the polynomial terms in the instanton partition function and isolate the coupling dependence that flows into $A$ and $B$ from the one affecting the non-perturbative terms. This will allow us to push the computation of $A$ and $B$ up to $4$ instantons in Section \ref{Sec:computation}. In Section \ref{Sec:comparison} we will show that our results perfectly agree with those of \cite{Hellerman:2020sqj,Hellerman:2021yqz}, which have been derived with different methods. Finally, in Section \ref{Sec:AllOrd}, we will prove at arbitrary instanton order (i.e.~exactly in $\tau$) that the large-vev sector, consisting solely of the contributions of $\mathcal{F}_0$ and $\mathcal{F}_1$ to the free energy, not only reproduces the correct coefficients of the $n\log n$ and $\log n$ terms, in agreement with \cite{Grassi:2019txd}, but also affects only the $\tau$ dependence of the coefficient $A(\tau,\bar{\tau})$ in the large-charge expansion \eqref{eq:log_corr}.

\subsection{Four-instanton calculation}\label{Sec:computation}

The polynomial truncation of the instanton partition function, which amounts to the instanton corrections to $\mathcal{F}_0$ and $\mathcal{F}_1$ only, reads
\be
    Z_{\text{inst}}({\rm i}a,\tau) &=& 1 + \frac{q}{2}\,(a^2-3) + \frac{q^2}{4}\,\left(\frac{a^4}{2}-\frac{35}{16} a^2 + \frac{13}{8}\right) + \frac{q^3}{8}\,\left(\frac{a^6}{6} - \frac{11}{16}a^4 + \frac{7}{48} a^2 + \frac{3}{8} \right) \nonumber \\
    && + \frac{q^4}{16}\, \left(\frac{a^8}{24} - \frac{3}{32} a^6 - \frac{517}{1536} a^4 +\frac{285}{2048}a^2 + \frac{141}{512} \right)+\mathcal{O}(q^5)\,.
\ee

We follow the same steps as in Section \ref{Sec:pol}. The $3$- and $4$-instanton contributions to the holomorphic matrix $\mathcal{M}(\tau)$ in \eqref{eq:detcalM} are respectively
\be
\mathcal{Z}_{ij}^{(3)}&=&(a^2_j+3)^i\left(\frac{a^6_j}{6}-\frac{11}{16} a^4_j + \frac{7}{48} a^2_j + \frac{3}{8}\right)\,, \nonumber \\
\mathcal{Z}_{ij}^{(4)}&=&(a^2_j+4)^i\left(\frac{a^8_j}{24}-\frac{3}{32} a^6_j - \frac{517}{1536} a^4_j + \frac{285}{2048} a^2_j + \frac{141}{512}\right)\,,
\ee
and its determinant reads
\begin{align}\label{Eq:HolM4}
\det\mathcal{M}(\tau)\;=\;&1+\frac{q}{2}\,\Tr\,\mathcal{K}^{(1)}+\frac{q^2}{4}\,\left[\Tr\left(\mathcal{K}^{(2)}-\frac{\left(\mathcal{K}^{(1)}\right)^2}{2}\right)+\frac{\left(\Tr\,\mathcal{K}^{(1)}\right)^2}{2}\right]\nonumber\\ 
&+\frac{q^3}{8}\,\left[\Tr\left(\mathcal{K}^{(3)}-\mathcal{K}^{(1)}\mathcal{K}^{(2)}+\frac{\left(\mathcal{K}^{(1)}\right)^3}{2}\right)\right.\nonumber\\
&\left.+\Tr\,\mathcal{K}^{(1)}\Tr\,\mathcal{K}^{(2)}-\frac{\Tr\,\mathcal{K}^{(1)}\,\Tr\left(\mathcal{K}^{(1)}\right)^2}{2}+\frac{\left(\Tr\,\mathcal{K}^{(1)}\right)^3}{6}\right]
\nonumber\\ &+\frac{q^4}{16}\,\left[\Tr\left(\mathcal{K}^{(4)}-\frac{\left(\mathcal{K}^{(2)}\right)^2}{2}-\mathcal{K}^{(1)}\mathcal{K}^{(3)}+\left(\mathcal{K}^{(1)}\right)^2\mathcal{K}^{(2)}-\frac{\left(\mathcal{K}^{(1)}\right)^4}{4}\right)\right.\nonumber\\
&+ \left.\frac{\left(\Tr\left(\mathcal{K}^{(1)}\right)^2\right)^2}{8}+\frac{\Tr\,\mathcal{K}^{(1)}\,\Tr\left(\mathcal{K}^{(1)}\right)^3}{3}-\Tr\,\mathcal{K}^{(1)}\,\Tr\left(\mathcal{K}^{(1)}\mathcal{K}^{(2)}\right)\right.\nonumber\\
&+\left.\Tr\,\mathcal{K}^{(1)}\Tr\,\mathcal{K}^{(3)}+\frac{\left(\Tr\,\mathcal{K}^{(2)}\right)^2-\Tr\,\mathcal{K}^{(2)}\Tr\left(\mathcal{K}^{(1)}\right)^2}{2}\right.\nonumber\\
&+ \left.\frac{\Tr\,\mathcal{K}^{(2)}\left(\Tr\,\mathcal{K}^{(1)}\right)^2}{2}-\frac{\Tr\left(\mathcal{K}^{(1)}\right)^2\left(\Tr\,\mathcal{K}^{(1)}\right)^2}{4}+\frac{\left(\Tr\,\mathcal{K}^{(1)}\right)^4}{24}\right]\nonumber\\
&+\mathcal{O}(q^5)\,.
\end{align}
We have already studied the first line of \eqref{Eq:HolM4} in Section \ref{Sec:pol}. Hence we proceed to discuss the cubic and the fourth orders. It is straightforward to verify that for the term of order $q^3$ we have
\begin{align}\label{orderq3}
\frac{q^3}{8}\;\cdot&\left[\frac{(2n^3-22n^2+67n-18)(n^2-5n-2)(n+1)}{96}+\frac{12 n^4-120 n^3 + 249 n^2 + 303 n + 14}{96} \sum_{\alpha=0}^n a^2_\alpha\right.\nonumber\\
& \left.  +\frac{1}{4} \left(\frac{13}{4} + (n-6)(n+1)\right) \left(\sum_{\alpha=0}^n a^2_\alpha\right)^2 + \frac{1}{6}\left(\sum_{\alpha=0}^n a^2_\alpha\right)^3 \right]\,,
\end{align}
while for the term of order $q^4$ we have
\begin{align} \label{orderq4}
\frac{q^4}{16}\;\cdot&\left[\frac{(n+1)(8 n^7 - 168n^6 + 1392 n^5 - 5560 n^4 + 10174 n^3 -5618 n^2 - 1269 n+846)}{3072}\right.\nonumber\\
&\left.+\frac{128 n^6 - 1920 n^5 + 9648 n^4 - 14944 n^3 - 8729 n^2 + 10664 n + 855}{6144} \sum_{\alpha=0}^n a^2_\alpha \right.\nonumber\\
&\left.+ \frac{1}{24}\left(\frac{3}{2}n^4 - 15 n^3 + 36 n^2 + \frac{27}{2} n - \frac{517}{64}\right)\,\left(\sum_{\alpha=0}^n a^2_\alpha\right)^2\right.\nonumber\\
&\left.+\frac{1}{24}\left(2 n^2 - 10 n - \frac{9}{4}\right)\,\left(\sum_{\alpha=0}^n a^2_\alpha\right)^3 + \frac{1}{24}\left(\sum_{\alpha=0}^n a^2_\alpha\right)^4\right]\,.
\end{align}

Jumping directly to the logarithm of $\det M^{(n)} (\tau,\bar\tau)$, we of course re-obtain Eq.~\eqref{Eq:detMCum} for the first two orders, whereas for the third order we find 
\be\label{eq:order3}
\log \det M^{(n)} (\tau,\bar\tau)|_{q^3}&=& \frac{q^3+\overline{q}^3}{192}\left[(19 n-90)(n+1) + 23 \langle \Tr\,\chi\rangle_c +  \frac{39}{2} \langle (\Tr\,\chi)^2\rangle_c\right] \nonumber\\ &&+ \frac{13(q^2 \overline{q} + q \overline{q}^2)}{128}   \langle (\Tr\,\chi)^2 \rangle_c + \frac{(q+\overline{q})^3}{48} \langle (\Tr\,\chi)^3 \rangle_c\,, 
\ee
while for the fourth order we get
\be\label{eq:order4}
\log \det M^{(n)} (\tau,\bar\tau)|_{q^4}&=& \frac{q^4+\overline{q}^4}{1536} \Bigl[\frac{3(1257n-5678)(n+1)}{32} + \frac{8103}{64}\langle \Tr\,\chi\rangle_c + \frac{1979}{16}\langle (\Tr\,\chi)^2\rangle_c \nonumber\\
&&  + 39 \langle (\Tr\,\chi)^3\rangle_c \Bigr]+ \frac{q^3 \overline{q} + q \overline{q}^3}{384} \left(23\langle (\Tr\,\chi)^2 \rangle_c + \frac{39}{2} \langle (\Tr\,\chi)^3 \rangle_c\right)\nonumber\\
&&+ \frac{q^2 \overline{q}^2}{256} \left(\frac{169}{16}\langle (\Tr\,\chi)^2 \rangle_c + 13 \langle (\Tr\,\chi)^3 \rangle_c\right)+\frac{(q+\overline{q})^4}{384} \langle (\Tr\,\chi)^4\rangle_c\,.
\ee
The above formulas confirm a key fact that we anticipated in Section \ref{Sec:pol}: Only the \emph{cumulants} appear in the final result. Here, in particular, in addition to the one- and two-point cumulants, we also need the three- and four-point cumulants, which are defined, respectively, as
\begin{align}
\langle (\Tr\,\chi)^3\rangle_c = & \,\langle (\Tr\,\chi)^3 \rangle - 3 \langle (\Tr\,\chi)^2 \rangle \langle \Tr\,\chi \rangle + 2 \langle \Tr\,\chi\rangle^3 \,,\nonumber\\
 \langle (\Tr\,\chi)^4\rangle_c =&\, \langle (\Tr\,\chi)^4 \rangle - 4 \langle (\Tr\,\chi)^3 \rangle \langle \Tr\,\chi\rangle + 12 \langle (\Tr\,\chi)^2 \rangle \langle \Tr\,\chi\rangle^2 - 3 \langle (\Tr\,\chi)^2\rangle^2 - 6 \langle \Tr\,\chi \rangle^4\,.
\end{align}
Recall that only the specific combination of expectation values appearing in the cumulants guarantees that, no matter how high the power of $\Tr\,\chi$ is, the power of $n$ that survives in the final result is at most $2$, compatibly with the EFT predictions. It is indeed straightforward to verify, using \eqref{eq:Yn_explicit_pert} and the definition of cumulants, that
\be\label{eq:cumgen}
\langle (\Tr\,\chi)^p\rangle_c &\underset{n \to \infty}{\simeq}& (p-1)!\,\frac{(n+1)(n+2)}{t^p}\qquad\; p>0\,,
\ee
where we recall that $t\equiv  4 \pi \text{Im}\tau+8\log2$. This phenomenon, which at this stage may appear somewhat miraculous, will find a natural explanation in our general argument presented in Section \ref{Sec:AllOrd}.
Finally, using \eqref{eq:corr_def}, \eqref{Eq:detMCum}, \eqref{eq:order3}, \eqref{eq:order4}, and \eqref{eq:cumgen}, we can compute the full $4$-instanton contribution to the perturbative large-$n$ expansion of the logarithm of the ($Z_{S^4}$-rescaled) two-point function:
\be\label{eq:log_corr_4inst}
\log(Z_{S^4}(\tau,\bar{\tau}) G_{2n}(\tau,\bar{\tau})) &\underset{n \to \infty}{\simeq} &\log(Z_{S^4}(\tau,\bar{\tau}) G_{2n}(\tau,\bar{\tau}))_0+ 2 (n+1) \sum_{k=1}^4 \frac{1}{k} \left(\frac{q+\overline{q}}{2 t}\right)^k\nonumber\\
&+&\hspace{-.5cm}\, \frac{q+\bar{q}}{2} \left(n-3\right) \nonumber \\
&+&\hspace{-.5cm}\,\frac{q^2+\bar{q}^2}{32} \left(9 n - 23+(n+1)\frac{13}{t}\right)\nonumber\\
&+&\hspace{-.5cm}\,\frac{q^3+\overline{q}^3}{96}\left(19 n - 45 + (n+1)\left(\frac{23}{t} +  \frac{39}{2t^2}\right)\right)\nonumber\\
&+&\hspace{-.5cm}\,\frac{q^4+\overline{q}^4}{8192} \left(1257 n - 2839 + (n+1)\left(\frac{2701}{2t} + \frac{3958}{3t^2} + \frac{832}{t^3} \right)\right)\nonumber\\
&+&\hspace{-.5cm}\,q \overline{q}\frac{(n+1)}{64}\left[(q + \overline{q})\frac{13}{t^2}+(q^2 +\overline{q}^2) \left(\frac{23}{3t^2} + \frac{13}{t^3} \right) + q \overline{q} \left(\frac{169}{32t^2} + \frac{13}{t^3}\right)\right]\nonumber\\
&+&\hspace{-.5cm}\,\mathcal{O}(q^5)\,,
\ee
where the result at $0$ instantons, which coincides with what was found in \cite{Grassi:2019txd}, is reported in Eq.~\eqref{eq:0instlogG}.

This final expression clearly shows that instanton contributions to the large-$n$ asymptotic expansion of the logarithm of the correlator \eqref{2ptF} are either exponentially suppressed in $n$, or limited to the sole coefficients of $n^1$ and $n^0$. Such a structure persists at any instanton order, as we will show in closed form in Section \ref{Sec:AllOrd}, and it is in perfect agreement with the expectations of the EFT, Eq.~\eqref{eq:log_corr}.

\subsection{The coefficients $A$ and $B$}\label{Sec:comparison}

In the final result of the previous Section, Eq.~\eqref{eq:log_corr_4inst}, we can extract the expressions, corrected up to $4$ instantons, for the coefficient $A$ (the term of order $n^1$) and for the coefficient $B$ (the scheme-dependent term of order $n^0$). In fact, the $\tau$-dependence of these coefficients is known exactly: It has been derived in \cite{Hellerman:2020sqj} and \cite{Hellerman:2021yqz} by employing different methods, such as S-duality and EFT techniques. In this Section, we will verify that these exact quantities, when expanded to the fourth order, reproduce our results.

\subsubsection*{$\bm{A(\tau,\bar\tau)}$}

The authors of \cite{Hellerman:2020sqj} derived the following exact expression for the coefficient $A$ as a function of the UV coupling $\tau$
\be\label{eq:exactA}
    A(\tau,\bar{\tau}) &=& \log F'(\tau) + \log\overline{F'(\tau)}-\log\left[-4 \left(F(\tau)-\overline{F(\tau)}\right)^2\right]\,,
\ee
where $F(\tau)$ is the inverse of the modular lambda function, which gives the UV coupling in terms of the IR one, $q=\lambda(\tau_{\rm IR})$ \cite{Grimm:2007tm}:
\be\label{eq:F}
    F(\tau)& =& 2 \tau + \frac{4 i}{\pi} \log(2) - \frac{i}{\pi} \left( \frac{1}{2}q + \frac{13}{64} q^2 + \frac{23}{192} q^3 + \frac{2701}{32768} q^4 + \mathcal{O}(q^5)\right)\,.
\ee

We will provide an independent derivation of formula \eqref{eq:exactA} in Section \ref{Sec:AllOrd}. For now, let us perform its instanton expansion and compare it with the coefficient of $n^1$ in \eqref{eq:log_corr_4inst}. Denoting by $A^{(k)}(\tau,\bar\tau)$ the $k$th instanton contribution and recalling that $t\equiv  4 \pi \text{Im}\tau+8\log2$, we find the following
\be\label{Eq:As}
A^{(0)}(\tau,\bar\tau)&=&2\log\frac{\pi}{t}\,,\nonumber\\
A^{(1)}(\tau,\bar\tau)&=&\frac{q+\overline{q}}{2}\left(1+\frac{2}{t}\right)\,,\nonumber\\
A^{(2)}(\tau,\bar\tau)&=&\frac{q^2+\overline{q}^2}{32}\left(9+\frac{13}{t}\right)+\left(\frac{q+\overline{q}}{2t}\right)^2\,,\nonumber\\
A^{(3)}(\tau,\bar\tau)&=&\frac{q^3+\overline{q}^3}{96}\left(19+\frac{23}{t}+\frac{39}{2t^2}\right)+q\overline{q}(q+\overline{q})\frac{13}{64t^2}+\frac{2}{3}\left(\frac{q+\overline{q}}{2t}\right)^3\,,\nonumber\\
A^{(4)}(\tau,\bar\tau)&=&\frac{q^4+\overline{q}^4}{8192}\left(1257+\frac{2701}{2t}+\frac{3958}{3t^2}+\frac{832}{t^3}\right)+\frac{q\overline{q}(q^2+\overline{q}^2)}{64}\left(\frac{23}{3t^2}+\frac{13}{t^3}\right)\nonumber\\
&&+\,\frac{q^2\overline{q}^2}{64}\left(\frac{169}{32t^2}+\frac{13}{t^3}\right) +\frac{1}{2}\left(\frac{q+\overline{q}}{2t}\right)^4\,.
\ee
It is immediate to verify that these expressions are in perfect agreement with the results in Section \ref{Sec:computation}.\footnote{Note that, to extract the $0$-instanton contribution to $A(\tau,\bar\tau)$ from Eq.~\eqref{eq:log_corr_4inst}, we have to subtract from $\log(Z_{S^4}(\tau,\bar{\tau}) G_{2n}(\tau,\bar{\tau}))_0$ in Eq.~\eqref{eq:0instlogG} the quantity $4\log2+2\log2-2$, which comes from the explicit $4\log2$ and from the expansion of $\Gamma(2n+\tfrac52)$ in Eq.~\eqref{eq:log_corr}.}

\subsubsection*{$\bm{B(\tau,\bar\tau)}$}

For the coefficient of $n^0$, the authors of \cite{Hellerman:2021yqz} introduced the quantity $\tilde{B}(\tau,\bar\tau)=B(\tau,\bar\tau)-\log Z_{S^4}$, free of the counterterm ambiguities of the path integral that plague the partition function. Then they computed $B(\tau,\bar\tau)$ in a scheme where the $S^4$ partition function is the one used by AGT \cite{Alday:2009aq}, whose instanton contributions do not contain the $U(1)$ factor of the Nekrasov sum \cite{Nekrasov:2002qd}. Translated to the Pestun-Nekrasov scheme \cite{Pestun:2007rz}, their expression reads
\be\label{eq:exactB}
  B(\tau,\bar{\tau})\; = \; \tilde{B}(\tau,\bar{\tau})  + \log Z_{S^4} \; = \; 12 \log \gamma_G - 1 - \frac{9}{2} \log 2 - \frac{3}{2} \log \pi + \frac{1}{3}\left(\log q + \log \bar{q}\right)  \nonumber \\
    + \frac{4}{3} \left(\log(1-q) + \log(1-\bar{q})\right) -\, 2 \log\frac{F(\tau)-\overline{F(\tau)}}{2 i} - 4 \left(\log\eta (F(\tau))+\log\overline{\eta(F(\tau))}\right)\,,
\ee
with $Z_{S^4}$ as in Eq.~\eqref{eq:Z_S4}, $F(\tau)$ as in Eq.~\eqref{eq:F}, and $\eta$ the Dedekind eta function.

Let us now perform the instanton expansion of \eqref{eq:exactB} and compare it with the coefficient of $n^0$ in \eqref{eq:log_corr_4inst}. Denoting by $B^{(k)}(\tau,\bar\tau)$ the $k$th instanton contribution and recalling that $t\equiv  4 \pi \text{Im}\tau+8\log2$, we find the following
\be
B^{(0)}(\tau,\bar\tau)&=& 12 \log \gamma_G  + \frac{1}{6}\log 2 + \frac{1}{2}\log \pi - 2 \log t -1 \,,\nonumber\\
B^{(1)}(\tau,\bar\tau)&=&\frac{q+\overline{q}}{2}\left(-3 + \frac{2}{t}\right) \,,\nonumber\\
B^{(2)}(\tau,\bar\tau)&=&\frac{q^2+\overline{q}^2}{32}\left(-23+\frac{13}{t}\right)+\left(\frac{q+\overline{q}}{2t}\right)^2\,,\nonumber\\
B^{(3)}(\tau,\bar\tau)&=&\frac{q^3+\overline{q}^3}{96}\left(-45+\frac{23}{t}+\frac{39}{2t^2}\right)+q\overline{q}(q+\overline{q})\frac{13}{64t^2}+\frac{2}{3}\left(\frac{q+\overline{q}}{2t}\right)^3\,,\nonumber\\
B^{(4)}(\tau,\bar\tau)&=&\frac{q^4+\overline{q}^4}{8192}\left(-2839+\frac{2701}{2t}+\frac{3958}{3t^2}+\frac{832}{t^3}\right)+\frac{q\overline{q}(q^2+\overline{q}^2)}{64}\left(\frac{23}{3t^2}+\frac{13}{t^3}\right)\nonumber\\
&&+\,\frac{q^2\overline{q}^2}{64}\left(\frac{169}{32t^2}+\frac{13}{t^3}\right) +\frac{1}{2}\left(\frac{q+\overline{q}}{2t}\right)^4\,.
\ee
It is immediate to verify that these expressions are in perfect agreement with the results in Section \ref{Sec:computation}.\footnote{Note that, to extract the $0$-instanton contribution to $B(\tau,\bar\tau)$ from Eq.~\eqref{eq:log_corr_4inst}, we have to subtract from $\log(Z_{S^4}(\tau,\bar{\tau}) G_{2n}(\tau,\bar{\tau}))_0$ in Eq.~\eqref{eq:0instlogG} the quantity $\tfrac52\log2+\tfrac12\log\pi$, which comes from the expansion of $\Gamma(2n+\tfrac52)$ in Eq.~\eqref{eq:log_corr}.}

\subsection{All-order derivation}\label{Sec:AllOrd}

Motivated by the order-by-order analysis done in the previous Sections, we now truncate the free energy \eqref{Eq:FE} and just retain the terms $g=0$ and $g=1$, which are the leading ones in the large-vev regime. What makes this truncation particularly useful is that exact expressions for both $\mathcal{F}_0$ and $\mathcal{F}_1$ are known. Starting from $\mathcal{F}_0$, recall that it can be written as \cite{Seiberg:1994aj}
\be\label{Eq.F0}
\mathcal{F}_0&=& \frac{1}{2}\,\tau_{\rm IR}\,a^2\;=\;\frac{1}{2}\,F(\tau)\,a^2\,,
\ee
where the function $F(\tau)$, whose weak-coupling expansion is displayed in Eq.~\eqref{eq:F}, maps the UV gauge coupling to the effective one in the IR \cite{Grimm:2007tm}. This function encodes the only data of the Seiberg-Witten theory of $SU(2)$ SQCD with four massless fundamental hypers which cannot be fixed by dimensional analysis. From Eq.~\eqref{Eq.F0}, using Matone's relation \cite{Matone:1995rx}, we can extract the exact expression for the vev of the chiral-ring generator \eqref{Eq:CBop}, which parametrizes the family of elliptic curves in the Seiberg-Witten theory:
\be\label{Eq:Matone}
u\;\equiv\;2\langle\Tr\phi^2\rangle&=&\frac{\partial\mathcal{F}_0}{\partial\tau}\;=\;\frac{1}{2}\,F'(\tau)\,a^2\,.
\ee

As for $\mathcal{F}_1$, its exact expression can be extracted from the geometry of the Seiberg-Witten curve \cite{Moore:1997pc} (see also \cite{Manschot:2019pog}):
\be
\mathcal{F}_1&=&\frac{1}{12}\,\log\Delta(u)-\frac{1}{2}\,\log\frac{\partial u}{\partial a}(u)\;=\;\frac{3}{2}\log a+\log F'(\tau)+{\rm const}\,, 
\ee
where $\Delta(u)\propto u^6$ is the physical discriminant of the Seiberg-Witten curve and, in the second equality, we used \eqref{Eq:Matone} and neglected irrelevant additive numerical constants.\footnote{These constants correspond to rescaling the partition function by pure numbers, which does not affect any scheme-independent quantity, and hence any observable.}

With the above quantities we can write the truncated version of the four-sphere partition function, which takes the form
\be\label{Eq:TruncZ}
Z_{S^4}|_{g=0,1}&=&\int_\mathbb{R}{\rm d}a\,\left|e^{2\pi{\rm i}\mathcal{F}_0}\right|^2\,\left|e^{\mathcal{F}_1}\right|^2\;\propto\;\frac{\left|F'(\tau)\right|^2}{\left(F(\tau)-\overline{F(\tau)}\right)^2}\;=\;\partial_\tau\partial_{\bar\tau}\log\left(F(\tau)-\overline{F(\tau)}\right)\,,
\ee
where again, in the second step, we neglected irrelevant multiplicative numerical constants. Note that the large-vev truncation crucially makes the integral in $a$ trivial to perform, yielding simply a function of $\tau$. The latter, in turn, can be interpreted geometrically as the pull-back through $F(\tau)$ of the hyperbolic metric on the IR upper-half-plane, whose K\"ahler potential is $\log{\rm Im}F(\tau)$, as shown in the last equality. This peculiarity translates into remarkable features of the corresponding matrix of derivatives \eqref{eq:matrixInt}. In particular, its determinant can be written as 
\be\label{Eq:DetMTrunc}
\det M^{(n)}(\tau,\bar\tau)|_{g=0,1} &=& a_{n+1}\left(\frac{\left|F'(\tau)\right|}{F(\tau)-\overline{F(\tau)}}\right)^{n(n+1)}\,,
\ee
where $a_{n+1}$ is a number satisfying the following recursion relation
\be\label{Eq:recurs}
a_{n+1}&=&(-1)^n\,n!\,(n+1)!\,a_n\,,\qquad \quad a_1\equiv1\,.
\ee
Equation \eqref{Eq:DetMTrunc} shows that, in computing the determinant, all derivatives of $F(\tau)$ higher than the first one drop out! Therefore, the result ends up being proportional to a power of the partition function itself. Moreover, such a power is only quadratic in $n$, because it originates from taking $\sum_i^ni$ holomorphic derivatives of the K\"ahler potential and just as many anti-holomorphic ones. This explains why in our order-by-order analysis all powers of $n$ higher than two cancel out in the cumulants.\footnote{Recall that, from the matrix-model perspective, each $t$-derivative produces two powers of $n$, as is manifest from Equations \eqref{Eq:DerX0} and \eqref{eq:Yn_explicit_pert}.}

Using Equations \eqref{eq:corr_def}, \eqref{Eq:DetMTrunc}, and \eqref{Eq:recurs}, the logarithm of the two-point function \eqref{2ptF} reads
\be\label{Eq:G2nExact}
\log G_{2n}(\tau,\bar{\tau}) &=& \log\left(\Gamma(n+1) \Gamma(n+2)\right)\nonumber\\&+&n\left\{4\log2+\log F'(\tau) + \log\overline{F'(\tau)}-\log\left[-\left(F(\tau)-\overline{F(\tau)}\right)^2\right]\right\}\,.
\ee
Note that this expression is also exact in $n$, because here, as opposed to the previous Sections (see e.g.~\eqref{eq:log_corr_4inst}), we did not include the full one-loop partition function, but only its large-vev portion. Comparing \eqref{Eq:G2nExact} with the large-charge expansion \eqref{eq:log_corr}, which defines the coefficients $A(\tau,\bar\tau)$ and $\tilde{B}(\tau,\bar\tau)$, we see that the former coincides with the exact expression \eqref{eq:exactA} found in \cite{Hellerman:2020sqj}, whereas the latter turns out to be independent of $\tau$:
\be\label{Eq:TildeBexact}
\tilde{B}(\tau,\bar\tau)|_{g=0,1}&=&\frac{1}{2}\log\frac{\pi}{8}\,.
\ee
It is important to remark, however, that, while $A(\tau,\bar\tau)$ receives no contributions from small vevs, Eq.~\eqref{Eq:TildeBexact} only gives the large-vev piece of $\tilde{B}(\tau,\bar\tau)$. Thus, we conclude that the non-trivial $\tau$ dependence of $\tilde{B}(\tau,\bar\tau)$ originates exclusively from the small-vev sector of the partition function.

\section{Outlook on Argyres-Douglas theories}\label{Sec:Concl}

In this note, we examined how the large-R-charge asymptotic expansion of the two-point function of Coulomb-branch operators depends on the exactly-marginal coupling in four-dimensional $\mathcal{N}=2$ $SU(2)$ SQCD with four massless fundamental flavors. We used the localization method of \cite{Gerchkovitz:2016gxx} to compute the correlator \eqref{2ptF}, and showed that, for a large number of insertions, its logarithm behaves as in formula \eqref{eq:log_corr}, thus providing the first analytic test of the EFT predictions of \cite{Hellerman:2018xpi}. In particular, we extended the one-loop analysis of \cite{Grassi:2019txd} by including the instanton partition function, which is not negligible if the large-charge limit is performed at fixed Yang-Mills coupling. Despite the fact that our fully general study was limited to two instantons, the strategy we adopted can in principle be carried over up to any instanton order, and we claim it to yield the same conclusions. We concluded that, at the instanton level, the perturbative large-charge expansion of the correlator is affected only by the leading terms in the large-radius (or large-vev) expansion of the free energy, i.e.~$\mathcal{F}_0$ and $\mathcal{F}_1$, which, moreover, contribute uniquely to the coefficients of order $n^1$ and $n^0$. All other instanton effects are exponentially suppressed in the large-$n$ limit. This is in striking contrast to the one-loop level, where, as shown in \cite{Grassi:2019txd}, the small-radius behavior of the free energy is crucial to reproduce the correct asymptotic expansion of the correlator.

We observe that the absence of instanton contributions in the perturbative $1/n$ expansion is a distinctive feature also exhibited by the large-charge limit of (maximal-trace) integrated four-point correlators of superconformal primary operators in $\mathcal{N}=4$ $SU(2)$ Super-Yang-Mills \cite{Brown:2023why,Paul:2023rka}.\footnote{We thank Congkao Wen for drawing our attention to this point.} These observables provide the closest analog, within the $\mathcal{N}=4$ setting, to extremal two-point functions in $\mathcal{N}=2$ SCFTs. However, this property does not persist at higher rank: In the $\mathcal{N}=4$ theory with gauge group $SU(N)$, the coefficients in the $1/n$ expansion acquire a non-trivial $\tau$ dependence through non-holomorphic Eisenstein series \cite{Brown:2023why,Paul:2023rka}. By analogy, one is therefore led to expect that instantons should contribute to the negative powers of $n$ in the large-charge expansion of $\log G_{2n}(\tau,\bar\tau)$ for $\mathcal{N}=2$ $SU(N)$ SQCD with $N>2$. Although such a contribution would be broadly consistent with the EFT perspective, due to the emergence of higher-derivative F-terms at higher rank \cite{Hellerman:2018xpi}, it would nevertheless be valuable to confirm this expectation through a direct localization computation along the lines of the one presented here.\\

Given the very limited role that instantons play at large charge in the scale-invariant $SU(2)$ SQCD, we would be tempted to speculate that the same holds in the asymptotically-free cases (one, two, and three fundamental flavors), which, at special values in their (extended) Coulomb branch, are known to flow to the three rank-one Argyres-Douglas SCFTs (see e.g.~\cite{Argyres:1995xn}). If this were literally the case, the one-loop Nekrasov partition function of the gauge theory would be sufficient to read off the universal behavior at large charge of the correlator of the corresponding Argyres-Douglas theory, Eq.~\eqref{Eq:GenHO}, which, expressing $\alpha$ in terms of $d$, becomes\footnote{Recall that, in the context of gauge theory, $d$ emerges as the rational number controlling the $a$ dependence of the vev of the Coulomb-branch operator \eqref{Eq:CBop} around the Argyres-Douglas point $a_*$, i.e.~$\langle\mathcal{O}\rangle\sim(a-a_*)^d$.}
\be\label{Eq:G2nAD}
G_{2n}|_{\rm universal}&\underset{n \to \infty}{\simeq}& \Gamma\left(dn+\frac{3d-1}{2}\right)\,.
\ee
 
However, such a circumstance is unlikely to happen because, as opposed to the conformal case, gauge-theory instantons are decisive to determine how the large-radius approximation of the free energy, and in particular $\mathcal{F}_1$, depends on $a$ at strong coupling \cite{Moore:1997pc}.\footnote{For example, in the case of one flavor, $\mathcal{F}_1$ tends to a constant in the weakly-coupled region $a\to\infty$, whereas it goes like $\tfrac3{10}\log(a-a_*)$ around the Argyres-Douglas point $a_*$. In the conformal case, instead, the instanton sum only contributes an $a$-independent addend in $\mathcal{F}_1$.} The large-radius free energy, in turn, enters prominently the leading behavior of the correlator at large charge. Indeed,  by truncating the free energy to just $\mathcal{F}_0$ and $\mathcal{F}_1$, analogously to what we did in Section \ref{Sec:AllOrd},\footnote{For $d=2$, we are actually neglecting instantons and focusing only on the tree-level and one-loop parts of $\mathcal{F}_0$ and $\mathcal{F}_1$. In this case, taking derivatives w.r.t.~the coupling corresponds to simply inserting powers of the Coulomb-branch operator, whose normalization here we are careless about.} one gets for the matrix \eqref{eq:matrixInt} \cite{Bissi:2021rei}
\be\label{Eq:MnAD}
M^{(n)}_{i j}&\propto&\Gamma\left(\frac{d}{2}(i+j+3)-1\right)\,,
\ee
where we neglected all numerical factors in front, as well as the normalization by the partition function, which do not alter the universal behavior of the correlator at large charge. It is possible to show, numerically, that the ratio of determinants \eqref{eq:corr_def}, with $M^{(n)}$ given by \eqref{Eq:MnAD}, has the same leading universal behavior of \eqref{Eq:G2nAD}, i.e.~Exp$[(dn+\tfrac32d-1)\log n]$, while it fails to reproduce the subleading terms $\mathcal{O}(e^{1/n})$.

The situation at small radius is radically different. Here, the instanton partition function can be organized as a power series in the small quantity $\Lambda R$, where $\Lambda$ is the dynamically-generated scale of the gauge theory and $R$ is the sphere radius
\be
Z_{\rm inst}&=&1+\sum_{k=1}^\infty a_k(a,m,R) (\Lambda R)^k\,,
\ee
where $a_k(a,m,R)$ are dimensionless functions, regular in the limit $R\to0$. As a consequence, there is no instanton contribution to the leading $R^0$ order.\footnote{Instead, in the conformal case, the instanton sum yields again an $a$-independent function of $\tau$ at order $R^0$, which however, as we learned in Section \ref{Sec:RoleInst}, produces only exponentially-suppressed contributions in the large-$n$ $\log(Z_{S^4} G_{2n})$.} Concerning the one-loop partition function, it turns out that all three asymptotically-free gauge theories are such that $Z_{\rm 1L}=1+\mathcal{O}(R^2)$ for $R\to0$. This implies that, around the Argyres-Douglas point, the reduced partition function of the gauge theory $\hat{Z}\equiv e^{-\mathcal{F}_0 R^2}Z$ goes to a constant in the limit $R\to0$.\footnote{This can be independently confirmed by analyzing the partition function of Argyres-Douglas theories via the holomorphic anomaly equation \cite{Fucito:2023txg}.}

The above discussion leads us to an educated guess for a function \emph{interpolating} between the large-radius and the small-radius regimes of the partition function
\be\label{Eq:Guess}
|\hat{Z}(a)|^2|_{g=0,1,\infty}&\sim & (a^2 + {\rm const})^{\tfrac32(d-1)}\qquad\; d\neq2\,,
\ee
where $g=\infty$ indicates the leading contribution at small radius. In the same spirit as what was done in \cite{Grassi:2019txd} for the conformal case,\footnote{Note that in the conformal case, $d=2$, we have the Vandermonde term $a^2$ at small radius, which changes \eqref{Eq:Guess} to $|\hat{Z}(a)|^2|_{g=0,1,\infty}\sim a^2 \sqrt{a^2 + t}$ (see Eq.~\eqref{Eq:X_nCG}).} it would be very interesting to verify whether Eq.~\eqref{Eq:Guess} leads to two-point functions whose universal behavior at large charge fully matches the EFT prediction \eqref{Eq:G2nAD}. We expect such a behavior to be completely independent of the specific value of the constant appearing in Eq.~\eqref{Eq:Guess}. We leave this question to future study.\\

Regarding the non-universal behavior in the perturbative large-charge expansion, i.e.~the coefficients $A$ and $\tilde{B}$, we can actually use 
\eqref{Eq:MnAD} to compute them for other rank-one SCFTs, such as the three Argyres-Douglas theories \cite{Argyres:1995xn}, $\mathcal{H}_0,\mathcal{H}_1,\mathcal{H}_2$, for which $d=6/5,4/3,3/2$ respectively, and the three Minahan-Nemeschansky theories \cite{Minahan:1996cj}, $E_6,E_7,E_8$, for which $d=3,4,6$ respectively. All of them share the feature of being isolated models, namely they do not have exactly-marginal parameters. Therefore, for these SCFTs, the coefficients $A$ and $\tilde{B}$ are pure numbers. The results of the present paper for the SQCD case ($d=2$) leads us to argue that what we will find for the coefficient $A$ in the other SCFTs is the complete result (albeit dependent on the normalization of the Coulomb-branch operator), whereas we expect that the result for $\tilde{B}$ will be modified by the inclusion of small vevs.

To set up the computation, we take
\be
Z_{S^4}=\int_\mathbb{R}{\rm d}a \,e^{-a^2}a^{3(d-1)}\,,
\ee
where we used that $\mathcal{F}_1=\tfrac32(d-1)\log a$, modulo an irrelevant additive constant. Then we fix a convenient normalization of the Coulomb-branch operator, such that $u=a^d$, and consider the matrix\footnote{For $d=2$, this normalization corresponds to fixing the loop-renormalized coupling to the value $t=2\pi$.}
\be\label{Eq:MatricesAD}
M^{(n)}_{ij}&=&\frac{1}{Z_{S^4}}\int_\mathbb{R}{\rm d}a\, a^{d(i+j)}\,e^{-a^2}a^{3(d-1)}\;=\;\frac{\Gamma\left(\frac{d}{2}(i+j+3)-1\right)}{\Gamma\left(\frac{3d}{2}-1\right)}\,.
\ee
For $d\neq2$, to the best of our knowledge, the ratio of determinants \eqref{eq:corr_def} for the matrices \eqref{Eq:MatricesAD} can only be computed numerically. As mentioned above, such a ratio shares with the EFT prediction \eqref{Eq:G2nAD} the same coefficients of $n\log n$ and $\log n$, differing only for terms that vanish for $n\to\infty$. Therefore, we can compute $A$ using the following sequence\footnote{We ignore the factor $16^n$ in \eqref{eq:corr_def}.}
\be\label{eq:A_sequence}
   A = \lim_{n \to \infty} A_n\,, &&\qquad  A_n = \frac{\log (G_{2n}) - \log \Gamma \left(d n + \frac{3(d-1)}{2} \right)}{n}\,,
\ee
and, using the obtained value, we can compute $\tilde{B}$ using the following sequence

\be\label{eq:B_sequence}
   \tilde{B} = \lim_{n \to \infty} \tilde{B}_n\,, &&\qquad   \tilde{B}_n = \log (G_{2n}) - \log \Gamma \left(d n + \frac{3(d-1)}{2} \right)- A \, n \,.
\ee

A useful algorithm that accelerates the convergence is the so-called ``Richardson extrapolation'',\footnote{We thank Alba Grassi for suggesting this method.} which is defined as follows. Let $\{a_n\}_{n \ge 1}$ be a sequence, its Richardson extrapolation of order $N$ is
\be
    R_m^N[\{a_n\}_n]& =& \sum_{k=0}^N \frac{(-1)^{N+k} \, (k+m)^N \, a_{k+m}}{k! (N-k)!}\,.
\ee
Once numerical stability is reached to a chosen precision, by changing the parameters $(N,m)$, one approximates $\lim_{n}a_n$ with $R_m^N[\{a_n\}_n]$. Here we set $N=10$. Then we notice that, if we increase $d$, we should also increase $m$ in order to reach numerical stability. In particular, for the three Minahan-Nemeschansky models, the choice $m=200$ allows us to have stability for the first $9$ digits for the coefficient $A$ and $7$ digits for the coefficient $\tilde{B}$.\footnote{Note that the precision for $\tilde{B}$ is lower because it depends on the numerical value of $A$, which is itself subjected to an error.} In contrast, for the three Argyres-Douglas theories, choosing $m=10$ already ensures stability to the same level of approximation. In Table \ref{Tab:A_B} we report our approximation for $A$ and $\tilde{B}$ up to the $7$th decimal digit.
\begin{table}[H]
\centering
\begin{tabular}{|c|c|c|c|c|c|c|}
\hline
 & $\mathcal{H}_0$ & $\mathcal{H}_1$ & $\mathcal{H}_2$ & $E_6$ & $E_7$ & $E_8$ \\ \hline
$A$    & $-1.6904874$    & $-1.6298583$ & $-1.5605388$ & $-1.1373603$ & $-0.9662579$ & $-0.7444107$  \\ \hline
$\tilde{B}$    & $0.1507129$    & $0.1020099$  & $0.0044435$ & $-2.0144218$ & $-4.1600920$ & $-9.6973840$   \\ \hline
\end{tabular}
\caption{Values of $A$ and $\tilde{B}$ for the rank-one Argyres-Douglas and Minahan-Nemeschansky theories, extracted via the Richardson extrapolation.}\label{Tab:A_B}
\end{table}

As a crosscheck, we perform this numerical computation also for the SQCD case, i.e.~for $d=2$, where we have analytic expressions, recalling that \eqref{Eq:MatricesAD} corresponds to the regime of large vevs and perturbative in the gauge coupling. We find $A=-1.3862944$ and $\tilde{B}=-0.4673558$. The former approximates $A=-2\log2$, which is indeed the $0$-instanton value of $A$ displayed in \eqref{Eq:As}, if we set $t=2\pi$ to account for the peculiar normalization of the Coulomb-branch operator we have chosen here. The latter approximates $\tilde{B}=\tfrac12\log\tfrac\pi8$, which is indeed the coupling-independent, large-vev value of $\tilde{B}$ shown in \eqref{Eq:TildeBexact}.

\subsection*{Acknowledgements}

We are grateful to Alba Grassi and Cristoforo Iossa for their initial collaboration and for sharing many key insights with us. We also thank Massimo Bianchi, Agnese Bissi, Guillaume Dubach, Francesco Fucito, Francisco Morales, Domenico Orlando, and Congkao Wen for useful discussions and valuable comments on the manuscript. We finally acknowledge the Theory Division at CERN for kind hospitality during the early stage of this project.

\bibliographystyle{JHEP}
\bibliography{biblio.bib}

\end{document}